\documentclass[twocolumn,times]{aastex63}
\usepackage{jab,amsmath,comment,color}
\shorttitle{Solar Wind Turbulence Evolution}
\shortauthors{Chen et al.}

\newcommand{\vA}{\ensuremath{v_\mathrm{A}}}
\newcommand{\vrad}{\ensuremath{v_\mathrm{r}}}
\newcommand{\Ev}{\ensuremath{E_\mathrm{v}}}
\newcommand{\Eb}{\ensuremath{E_\mathrm{b}}}
\newcommand{\EB}{\ensuremath{E_\mathrm{B}}}
\newcommand{\Ep}{\ensuremath{E_\mathrm{+}}}
\newcommand{\Em}{\ensuremath{E_\mathrm{-}}}
\newcommand{\Et}{\ensuremath{E_\mathrm{t}}}
\newcommand{\fsc}{\ensuremath{f_\mathrm{sc}}}
\newcommand{\sigmac}{\ensuremath{\sigma_\mathrm{c}}}
\newcommand{\sigmar}{\ensuremath{\sigma_\mathrm{r}}}
\newcommand{\rA}{\ensuremath{r_\mathrm{A}}}
\newcommand{\rE}{\ensuremath{r_\mathrm{E}}}
\newcommand{\CB}{\ensuremath{C_\mathrm{B}}}

\newcommand{\RS}{\ensuremath{R_\odot}}
\newcommand{\MA}{\ensuremath{M_\mathrm{A}}}
\newcommand{\dzp}{\ensuremath{\delta\mathbf{z}^+}}
\newcommand{\dzm}{\ensuremath{\delta\mathbf{z}^-}}
\newcommand{\taub}{\ensuremath{\tau_\mathrm{b}}}
\newcommand{\tauc}{\ensuremath{\tau_\mathrm{c}}}
\newcommand{\kb}{\ensuremath{k_\mathrm{b}}}
\newcommand{\zp}{\ensuremath{\mathbf{z}^+}}
\newcommand{\zm}{\ensuremath{\mathbf{z}^-}}

\begin{document}

\title{The Evolution and Role of Solar Wind Turbulence in the Inner Heliosphere}
\author[0000-0003-4529-3620]{C. H. K. Chen}
\affil{School of Physics and Astronomy, Queen Mary University of London, London E1 4NS, UK}
\correspondingauthor{C. H. K. Chen}
\email{christopher.chen@qmul.ac.uk}
\author[0000-0002-1989-3596]{S. D. Bale}
\affil{Physics Department, University of California, Berkeley, CA 94720-7300, USA}
\affil{Space Sciences Laboratory, University of California, Berkeley, CA 94720-7450, USA}
\affil{The Blackett Laboratory, Imperial College London, London, SW7 2AZ, UK}
\affil{School of Physics and Astronomy, Queen Mary University of London, London E1 4NS, UK}
\author[0000-0002-0675-7907]{J. W. Bonnell}
\affil{Space Sciences Laboratory, University of California, Berkeley, CA 94720-7450, USA}
\author[0000-0002-0151-7437]{D. Borovikov}
\affil{Space Science Center, University of New Hampshire, Durham, NH 03824, USA}
\author[0000-0002-4625-3332]{T. A. Bowen}
\affil{Space Sciences Laboratory, University of California, Berkeley, CA 94720-7450, USA}
\author[0000-0002-8175-9056]{D. Burgess}
\affil{School of Physics and Astronomy, Queen Mary University of London, London E1 4NS, UK}
\author[0000-0002-3520-4041]{A. W. Case}
\affiliation{Smithsonian Astrophysical Observatory, Cambridge, MA 02138 USA}
\author[0000-0003-4177-3328]{B. D. G. Chandran}
\affil{Department of Physics \& Astronomy, University of New Hampshire, Durham, NH 03824, USA}
\affil{Space Science Center, University of New Hampshire, Durham, NH 03824, USA}
\author[0000-0002-4401-0943]{T. {Dudok de Wit}}
\affil{LPC2E, CNRS and University of Orl\'eans, Orl\'eans, France}
\author[0000-0003-0420-3633]{K. Goetz}
\affiliation{School of Physics and Astronomy, University of Minnesota, Minneapolis, MN 55455, USA}
\author[0000-0002-6938-0166]{P. R. Harvey}
\affil{Space Sciences Laboratory, University of California, Berkeley, CA 94720-7450, USA}
\author[0000-0002-7077-930X]{J. C. Kasper}
\affiliation{Climate and Space Sciences and Engineering, University of Michigan, Ann Arbor, MI 48109, USA}
\affiliation{Smithsonian Astrophysical Observatory, Cambridge, MA 02138 USA}
\author[0000-0001-6038-1923]{K. G. Klein}
\affiliation{Lunar and Planetary Laboratory, University of Arizona, Tucson, AZ 85719, USA}
\author[0000-0001-6095-2490]{K. E. Korreck}
\affiliation{Smithsonian Astrophysical Observatory, Cambridge, MA 02138 USA}
\author[0000-0001-5030-6030]{D. Larson}
\affil{Space Sciences Laboratory, University of California, Berkeley, CA 94720-7450, USA}
\author[0000-0002-0396-0547]{R. Livi}
\affil{Space Sciences Laboratory, University of California, Berkeley, CA 94720-7450, USA}
\author[0000-0003-3112-4201]{R. J. MacDowall}
\affil{Solar System Exploration Division, NASA/Goddard Space Flight Center, Greenbelt, MD 20771, USA}
\author[0000-0003-1191-1558]{D. M. Malaspina}
\affil{Laboratory for Atmospheric and Space Physics, University of Colorado, Boulder, CO 80303, USA}
\author[0000-0001-9202-1340]{A. Mallet}
\affil{Space Sciences Laboratory, University of California, Berkeley, CA 94720-7450, USA}
\author[0000-0001-6077-4145]{M. D. McManus}
\affil{Space Sciences Laboratory, University of California, Berkeley, CA 94720-7450, USA}
\author[0000-0002-9621-0365]{M. Moncuquet}
\affil{LESIA, Observatoire de Paris, Universit\'{e} PSL, CNRS, Sorbonne Universit\'{e}, Universit\'{e} de Paris, 92195 Meudon, France}
\author[0000-0002-1573-7457]{M. Pulupa}
\affil{Space Sciences Laboratory, University of California, Berkeley, CA 94720-7450, USA}
\author[0000-0002-7728-0085]{M. Stevens}
\affiliation{Smithsonian Astrophysical Observatory, Cambridge, MA 02138 USA}
\author[0000-0002-7287-5098]{P. Whittlesey}
\affil{Space Sciences Laboratory, University of California, Berkeley, CA 94720-7450, USA}

\begin{abstract}
The first two orbits of the \emph{Parker Solar Probe} (\emph{PSP}) spacecraft have enabled the first in situ measurements of the solar wind down to a heliocentric distance of 0.17\,au (or 36\,\RS). Here, we present an analysis of this data to study solar wind turbulence at 0.17\,au and its evolution out to 1\,au. While many features remain similar, key differences at 0.17\,au include: increased turbulence energy levels by more than an order of magnitude, a magnetic field spectral index of $-3/2$ matching that of the velocity and both Elsasser fields, a lower magnetic compressibility consistent with a smaller slow-mode kinetic energy fraction, and a much smaller outer scale that has had time for substantial nonlinear processing. There is also an overall increase in the dominance of outward-propagating Alfv\'enic fluctuations compared to inward-propagating ones, and the radial variation of the inward component is consistent with its generation by reflection from the large-scale gradient in Alfv\'en speed. The energy flux in this turbulence at 0.17\,au was found to be $\sim$10\% of that in the bulk solar wind kinetic energy, becoming $\sim$40\% when extrapolated to the Alfv\'en point, and both the fraction and rate of increase of this flux towards the Sun is consistent with turbulence-driven models in which the solar wind is powered by this flux.
\end{abstract}

\section{Introduction}

The solar wind is observed to contain a turbulent cascade at distances from the closest previous in situ measurements to the Sun at 0.29\,au \citep{tu95} out to the edge of the heliosphere and beyond \citep{fraternale19}. Our understanding of solar wind turbulence and the role it plays in the large scale dynamics, therefore, has come from measurements over this range of distances, much of which have been in the vicinity of 1\,au \citep{alexandrova13a,bruno13,kiyani15,chen16b}. The \emph{Parker Solar Probe} (\emph{PSP}) spacecraft \citep{fox16} has so far travelled nearly twice as close to the Sun, down to a heliocentric distance of 0.17\,au, and will get increasingly closer in future orbits. Measurements from \emph{PSP}, therefore, are allowing this new environment to be used to investigate the fundamental nature of plasma turbulence and the role it plays in the generation of the solar wind.

At 1\,au, it has long been known that the solar wind fluctuations at MHD scales are predominantly Alfv\'enic \citep{belcher71a,tu95,horbury05,bruno13} with a small energy fraction in compressive fluctuations that resemble the slow mode \citep{tu95,howes12a,klein12,bruno13,verscharen17}. The Alfv\'enic turbulence develops an anisotropic cascade that appears to be in critical balance \citep{horbury08,chen16b}, consistent with models of Alfv\'enic turbulence \citep{goldreich95,boldyrev06,lithwick07,beresnyak08,perez09,chandran15,mallet17}. However, the different MHD fields typically display different scalings, which depend on underlying parameters, such as the level of imbalance between the oppositely-directed Alfv\'enic fluxes, in a way that is not currently captured by any single model \citep{chen16b}. The compressive fluctuations are also highly anisotropic \citep{chen12b,chen16b} and thought to be passive with respect to the Alfv\'enic turbulence \citep{schekochihin09}. 

Previous missions, such as \emph{Helios}, \emph{Voyager}, and \emph{Ulysses} and the \emph{Mariner} spacecraft have allowed the radial evolution of the turbulence to be studied beyond 0.29\,au. Some key findings from this data have been decreasing power levels with increasing distance \citep{belcher74,villante80,bavassano82a,tu95,horbury01a}, a ``$1/f$'' break scale that moves to larger scales at greater distances \citep{bavassano82a,horbury96a,bruno13}, a correlation length that increases with distance \citep{tu95,ruiz14}, a reduction of the imbalance or cross-helicity with distance \citep{roberts87a,tu95,bavassano98,bavassano00,matthaeus04,breech05} and a velocity spectral index that evolves from $-3/2$ to $-5/3$ between 1 and 5\,au \citep{roberts10}. The evolution of all of these features is consistent with an active cascade occurring throughout the solar wind, which is also consistent with the observed non-adiabatic temperature profile suggesting continual heating of the plasma \citep{mihalov78,marsch82a,gazis82,freeman88,richardson95,matthaeus99a,cranmer09,hellinger11}.

In addition to this heating far from the Sun, turbulence is also proposed to play a key role in the heating of the solar corona and acceleration of the solar wind itself. Early solar wind models, based on the seminal work of \citet{parker58a}, were based on a thermally driven wind, but it was quickly reaslised that this was not sufficient to lead to observed solar wind properties 1\,au \citep[see reviews by][]{parker65,leer82,barnes92,hollweg08,hansteen12,cranmer15}. The propagation of Alfv\'en waves from the photosphere into the corona to drive a turbulent cascade was proposed as a possible solution; the waves and turbulence provide a pressure to directly accelerate the solar wind \citep{belcher71b,alazraki71} and the dissipation of the turbulence can provide additional heating \citep{coleman68}. The generation of this turbulence requires counter-propagating waves \citep{iroshnikov63,kraichnan65}, and the large-scale gradient in Alfv\'en speed was suggested to cause the outward-propagating waves to be partially reflected \citep{heinemann80,velli93} and initiate the cascade \citep{matthaeus99b,dmitruk02,cranmer05,verdini07,chandran09b,verdini09}. Modern turbulence-driven models now incorporate these components, together with other properties such as heat fluxes, pressure anisotropy, and turbulent dissipation, to achieve a self-consistent solar wind solutions that can match many properties of observational data \citep{cranmer07,verdini10,chandran11,vanderholst14,usmanov18}. However, the key test for these and other classes of solar wind model are measurements close to the Sun where the heating and acceleration are taking place.

In this paper, data from \emph{PSP} during its first two orbits is used to study turbulence down to a distance of 0.17\,au from the Sun for the first time. The basic properties of the turbulence are investigated, along with its radial evolution out to $\sim$1\,au, and compared to models of MHD turbulence and models of Alfv\'enic turbulence-driven solar wind, to determine the properties, evolution and role of solar wind turbulence in the inner heliosphere.

\section{Data}

The data used in this study, from the first two orbits of \emph{PSP} from 6th October 2018 to 18th April 2019, cover a heliocentric radial distance range 0.17 to 0.82\,au (or equivalently 35.7 to 174\,\RS). From the FIELDS instrument suite \citep{bale16}, magnetic field data, $\mathbf{B}$, from the outboard fluxgate magnetometer (MAG) averaged to 0.4369\,s resolution, and electron density, $n_\mathrm{e}$, derived from quasi-thermal noise (QTN) measurements made by the Radio Frequency Spectrometer Low Frequency Receiver (RFS/LFR) at 6.991\,s resolution \citep{moncuquet19}, were used. From the SWEAP instrument suite \citep{kasper16}, moments of the ion (proton) distributions (density $n$, velocity $\mathbf{v}$, and radial temperature $T_\mathrm{r}$) measured by the Solar Probe Cup (SPC) averaged to 27.96\,s resolution over the full orbit, and at a resolution of 0.8738\,s for a 1-day period during Perihelion 1, were used. In addition to the automated SPC data processing \citep{case19}, remaining unphysical data points were manually removed. Since the QTN density is more accurate closer to the Sun, a combination of QTN and SPC density was used: for each interval studied, the mean QTN density was used unless the density was not possible to calculate for more than half of the interval, in which case the average SPC density was used. This interval-averaged density was used to calculate the Alfv\'enic normalisation, plasma beta, and energy fluxes for the analysis in this paper.

The solar wind over the two orbits was mostly slow wind, Alfv\'enic in nature, with large amplitude ($\delta B/B\sim1$) fluctuations \citep{bale19,kasper19}. The orbits covered a mixture of source regions, although notably much of the first encounter was in wind from a small low-latitude coronal hole \citep{bale19,badman19}. The ratio of the solar wind speed to Alfv\'en speed, $v/\vA$, was always larger than 1 throughout both orbits, and larger than 3 the majority of the time, indicating the \citet{taylor38} hypothesis to be marginally well-satisfied which would enable temporal structure to be interpreted as spatial structure, i.e., spacecraft-frame frequencies \fsc\ to be interpreted as wavenumber $k$ through $k=(2\pi\fsc)/v$. However, even in those parts of the orbit where $v/\vA\sim1$ it has been shown that the Taylor hypothesis can hold for the dominant outward-propagating component of highly-imbalanced (i.e., high cross-helicity) turbulence \citep{klein15a}, and that when the Taylor hypothesis breaks down, the sweeping by larger-scale eddies leads to the same spectral index in the spacecraft-frame frequency spectra as the underlying wavenumber spectra \citep{bourouaine18,bourouaine19}. In this paper, the results are interpreted spatially.

\section{Results}

\subsection{Turbulence Spectrum}

\begin{figure}
\includegraphics[width=\columnwidth,trim=0 0 0 0,clip]{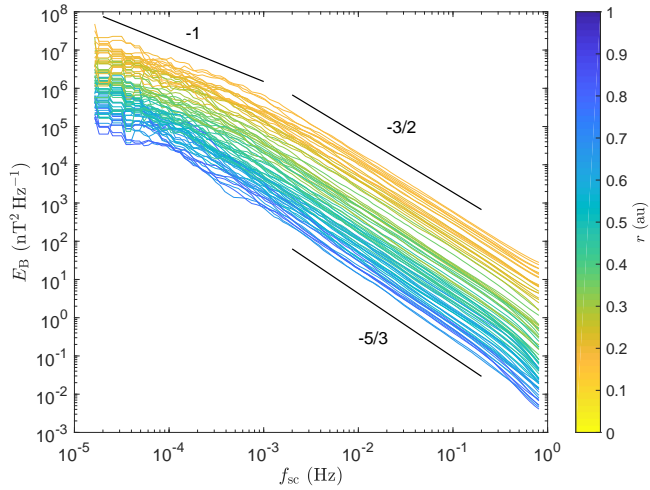}
\caption{Magnetic field power spectrum, \EB, at different heliocentric distances, $r$, over the first two \emph{PSP} orbits. Several power law slopes are marked for comparison. A turbulent inertial range is present at all distances, with a flattening at low frequencies. Deviations at high frequencies ($\fsc\gtrsim0.3$\,Hz) are partly due to digital filter effects.}
\label{fig:bspectrum}
\end{figure}

To examine the radial evolution of the magnetic field fluctuation spectrum, the MAG data were divided into one-day intervals for analysis. Periods containing coronal mass ejections were removed, all data gaps were linearly interpolated and days with more than 1\% of the data missing were excluded from the analysis. For each interval, the trace power spectral density was calculated by Fourier transform and, for clarity, smoothed by averaging over a sliding window of a factor of 2. The power spectra, \EB, as a function of spacecraft-frame frequency \fsc, are shown in Figure \ref{fig:bspectrum}, in which they are coloured by heliocentric distance, $r$. It can be seen that the power levels systematically increase as $r$ decreases by at least 2 orders of magnitude over the range considered. For frequencies $10^{-3}\,\mathrm{Hz}\lesssim\fsc\lesssim10^{-1}\,\mathrm{Hz}$, a power law range is present at all distances that is compatible with models of inertial range MHD turbulence (discussed below), and at lower frequencies a flattening, here compared to $\fsc^{-1}$, is present (although this low-frequency range is not the focus of the present study, see \citet{matteini19}). Typical ion kinetic scales are at $\fsc\gtrsim1$\,Hz \citep{duan19} so that all of the analysis in this paper corresponds to the MHD inertial range.

\begin{figure}
\includegraphics[width=\columnwidth,trim=0 0 0 0,clip]{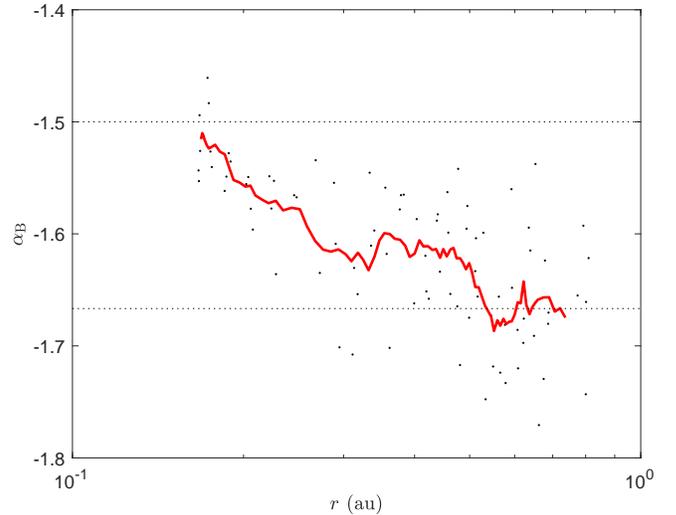}
\caption{Variation of magnetic field spectral index, $\alpha_\mathrm{B}$, with heliocentric distance, $r$, in the MHD inertial range ($10^{-2}\,\mathrm{Hz}<\fsc<10^{-1}\,\mathrm{Hz}$). The black dots show the spectral index measurements and the red line is a 10-point running mean. The horizontal dotted lines mark the theoretical predictions $-3/2$ and $-5/3$.}
\label{fig:bindex}
\end{figure}

A key diagnostic of the turbulence used to distinguish the nature of the cascade process is the power law spectral index $\alpha$, defined through $E\propto\fsc^\alpha$. This was calculated for each magnetic spectrum in the frequency range $10^{-2}\,\mathrm{Hz}<\fsc<10^{-1}\,\mathrm{Hz}$ and is shown as a function of radial distance in Figure \ref{fig:bindex}. A clear transition can be seen from $\alpha_\mathrm{B}\approx-3/2$ at $r\approx0.17$\,au to $\alpha_\mathrm{B}\approx-5/3$ at $r\approx0.6$\,au. This variation is consistent across all phases of the first two \emph{PSP} orbits and has not been observed before, since in situ measurements have previously only been available for $r\gtrsim0.3$\,au where the transition occurs. It can be seen that there is some scatter in the data; this may be in part due to statistical variation but could also be due to varying solar wind conditions and underlying parameters that control magnetic spectrum.

\begin{figure}
\includegraphics[width=\columnwidth,trim=0 0 0 0,clip]{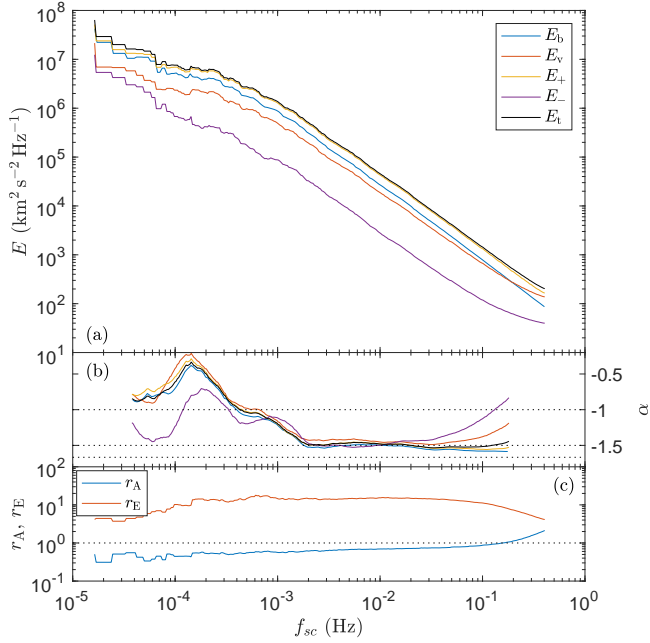}
\caption{(a) Spectra of Alfv\'enic turbulence variables at 0.17\,au. (b) Local spectral index $\alpha$ (calculated over a sliding window of a factor of 5), together with dotted lines marking values $-1$, $-3/2$, and $-5/3$. (c) Alfv\'en ratio, \rA, and Elsasser ratio, \rE.}
\label{fig:allspectra}
\end{figure}

Figure \ref{fig:allspectra} shows the trace spectra of the Alfv\'enic turbulence variables for the 24-hour period of the day of Perihelion 1, 6th November 2018, at 0.17\,au. The spectra are of the magnetic field in Alfv\'en units, $\mathbf{b}=\mathbf{B}/\sqrt{\mu_0\rho_0}$ where $\rho_0$ is the average mass density, the velocity $\mathbf{v}$, the \citet{elsasser50} variables, $\mathbf{z^\pm}=\mathbf{v}\pm\mathbf{b}$ describing the inward- and outward-propagating Alfv\'enic fluctuations, and the total energy $\Et=\Eb+\Ev=\Ep+\Em$ (note that the Elsasser spectra are defined with an additional factor of $\frac{1}{2}$ such that they sum to the total energy spectrum). It can be seen that all fields take a spectral index close to $\alpha\approx-3/2$ in the inertial range $\fsc\gtrsim2\times10^{-3}$\,Hz, until some (in particular \Em and \Ev) show an artificial flattening at high frequencies due to velocity noise\footnote{The $-3/2$ velocity spectrum extends down to the ion kinetic scales during the short periods when SPC was operating in flux angle mode, which has a lower noise level \citep{vech19}.}. This results in an approximately constant Alfv\'en ratio, $\rA=\Ev/\Eb$, and Elsasser ratio, $\rE=\Ep/\Em$, through the measured inertial range ($2\times10^{-3}\lesssim \fsc\lesssim5\times10^{-2}$\,Hz)\footnote{Note, however, that \citet{parashar19} report times in which the level of imbalance appears not to be constant through the inertial range.}. The average values, calculated as the mean of all of the values within this range, are $\rA=0.69$ and $\rE=14.6$, indicating highly-imbalanced outward-dominated Alfv\'enic turbulence with a small amount of residual energy\footnote{Pressure anisotropy can sometimes lead to significant modifications of the Alfv\'en ratio \citep{chen13b}, although were not found to be important here due to the low $\beta$.}.

\begin{figure}
\includegraphics[width=\columnwidth,trim=0 0 0 0,clip]{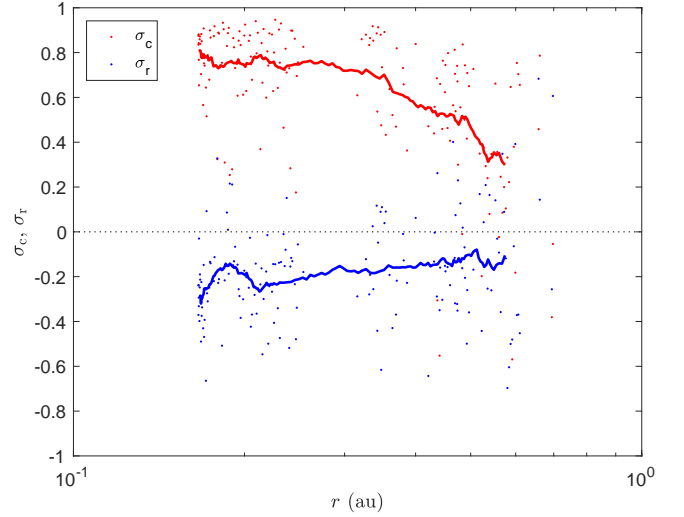}
\caption{Dependence of normalised cross-helicity, \sigmac, and residual energy, \sigmar, on heliocentric distance $r$. The dots mark 6-hour average values and the solid lines are 30-point running means.}
\label{fig:sigma}
\end{figure}

One possibility for the radial variation of the magnetic spectral index (Figure \ref{fig:bindex}) is that the shallower spectrum near the Sun reflects a transient stage of evolution, similar to the suggestion by \citet{roberts10} for the steepening of the velocity spectrum reported for $r>1$\,au. However, even by 0.17\,au there have been a large number of nonlinear times (see Section \ref{sec:outerscale}) meaning that the inertial range should already be in steady state by this distance. Another possibility is that the spectral index depends on an underlying parameter, such as the normalised cross-helicity $\sigmac=2\left<\delta\mathbf{b}\cdot\delta\mathbf{v}\right>/\left<\delta\mathbf{b}^2+\delta\mathbf{v}^2\right>$ or normalised residual energy $\sigmar=2\left<\delta\mathbf{z}^+\cdot\delta\mathbf{z}^-\right>/\left<\delta\mathbf{z}^{+2}+\delta\mathbf{z}^{-2}\right>$. Measurements at 1\,au \citep{podesta10d,wicks13b,chen13b,bowen18} have shown that $\alpha_\mathrm{B}$ depends on both of these quantities, taking a value of $\approx-3/2$ when $|\sigmac|\approx1$ or $|\sigmar|\approx0$ and steeper otherwise. To test this, the radial variation of \sigmac\ and \sigmar\ was calculated from 6-hour averages (with intervals containing heliospheric current sheet crossings \citep{szabo19} removed) and the results are shown in Figure \ref{fig:sigma}. The direction of $\mathbf{B}$ was ``rectified'' \citep{bruno85,roberts87a} with respect to the average sign of $B_\mathrm{r}$ over the interval so that $\mathbf{z}^+$ corresponds to outward-propagating Alfv\'enic propagation and $\mathbf{z}^-$ inwards. There is significant scatter, that reflects the varying solar wind conditions, but it can be seen that on average \sigmac\ decreases with increasing $r$ (from $\approx 0.8$ to $\approx 0.3$) and \sigmar\ is roughly constant at $\approx-0.2$\footnote{See \citet{mcmanus19} for details of the local properties of \sigmac\ and \sigmar\ measured by \emph{PSP} at perihelion.}. Therefore, the measurements are consistent with the previous dependence of $\alpha_\mathrm{B}$ on \sigmac\ at 1\,au, although this does not seem to be related to a change in residual energy.

Regarding the cause of the $-3/2$ spectra at 0.17\,au, this scaling is consistent\footnote{Since the local mean field is not being tracked, the measured frequency spectra can be interpreted as $k_\perp$ spectra, assuming $k_\perp\gg k_\|$ (as expected theoretically and measured a 1\,au \citep[e.g.,][]{chen16b}).} with models of both balanced \citep{boldyrev06,chandran15,mallet17} and imbalanced \citep{perez09,podesta10c} Alfv\'enic turbulence in homogeneous plasmas (e.g., without wave reflection) that involve scale-dependent alignment. In addition, recent simulations \citep{chandran19} of inhomogeneous reflection-driven MHD turbulence from the photosphere to 21\,\RS\ found that both $E_+$ and $E_-$ also tend towards $\alpha=-3/2$ past the Alfv\'en point for a range of values of the correlation time and perpendicular correlation length at the photosphere. \citet{chandran19} considered this in partial agreement with a reflection-driven version of the \citet{lithwick07} model of strong imbalanced MHD turbulence, which predicts the same spectral index ($\alpha=-5/3$) for both $E_+$ and $E_-$, with the $-3/2$ scaling possibly resulting from additional phenomena such as intermittency and scale-dependent alignment \citep[e.g.,][]{boldyrev06,chandran15}. However, it is also possible that the trend seen in Figure \ref{fig:bindex} is part way through a transition from an even shallower spectrum closer to the Sun or an effect of the driving on the cascade; these possibilities are discussed further in Section \ref{sec:discussion}.

\subsection{Magnetic Compressibility and Slow Mode Fraction}

\begin{figure}
\includegraphics[width=\columnwidth,trim=0 0 0 0,clip]{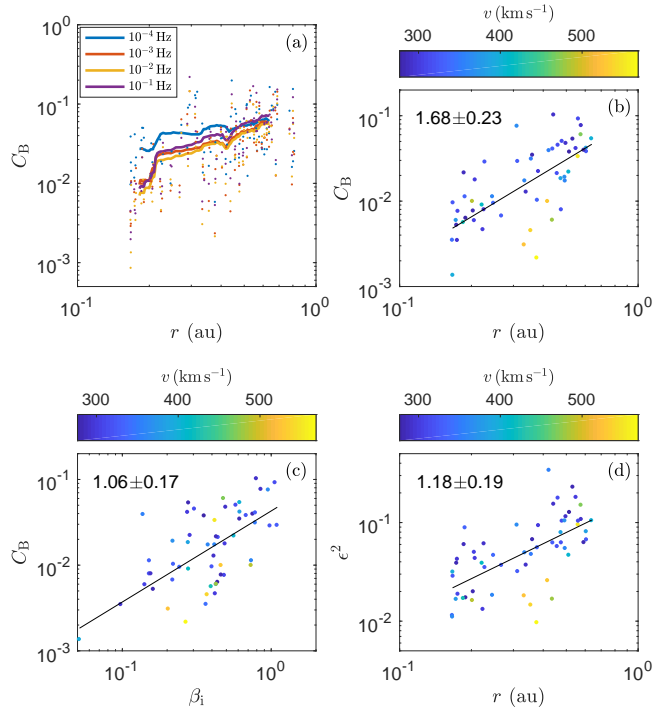}
\caption{(a) Magnetic compressibility, \CB, as a function of heliocentric distance, $r$, at four values of spacecraft-frame frequency \fsc; the solid lines are 30-point running means. (b) \CB\ as a function of $r$ at $\fsc=10^{-2}$\,Hz coloured by solar wind speed, $v$. (c) \CB\ as a function of ion plasma beta, $\beta_\mathrm{i}$. (d) Slow mode kinetic energy fraction, $\epsilon^2$, as a function of $r$.}
\label{fig:comp}
\end{figure}

Another well-known feature of solar wind turbulence is the low power in compressive fluctuations, in particular the low level of fluctuations in $|\mathbf{B}|$ \citep[e.g.,][]{bruno13,chen16b}. Figure \ref{fig:comp}(a) shows the magnetic compressibility, $\CB=\left(\delta|\mathbf{B}|/|\delta\mathbf{B}|\right)^2$, as a function of $r$ at four spacecraft-frame frequencies. There is a decrease of \CB\ towards smaller $r$ at all frequencies, which is independent of \fsc\ through the inertial range ($\fsc\gtrsim10^{-3}$\,Hz). Overall, the compressibility levels at perihelion are an order of magnitude smaller than at 1\,au. Figure \ref{fig:comp}(b) shows the compressibility as a function of $r$ at $\fsc=10^{-2}$\,Hz, coloured by solar wind speed. It can be seen that the periods of faster wind $v\gtrsim 500$\,km\,s$^{-1}$ (observed by \emph{PSP} on the outbound part of its first orbit between 0.3 and 0.5\,au) have a lower compressibility, consistent with previous observations at larger $r$ \citep{tu95,bruno13} and overall the data can be fit to a power law, $\CB\propto r^{1.68\pm0.23}$, although with significant scatter.

1\,au measurements \citep{howes12a,klein12,verscharen17}, as well as an analysis of \emph{PSP} data \citep{chaston19}, suggest that the compressive power is primarily in slow mode like fluctuations, so it is of interest to see if the radial variation in compressibility is due to varying $\beta$ or varying slow mode fraction. If it assumed that $\delta B_\perp$ arises from the Alfv\'en mode and $\delta |\mathbf{B}|\approx\delta B_\|$ from the slow mode, the compressibility is given by\footnote{While this is derived from MHD, which is not in principle applicable to the solar wind due to its low collisionality, recent measurements suggest the slow-mode fluctuations to be fluid-like in their polarisations \citep{verscharen17}, suggesting Equation (\ref{eq:comp}) may be a reasonable approximation.}
\begin{equation}
\CB=\frac{\epsilon^2\beta\gamma\sin^4(\theta_\mathrm{kB})}{2}
\label{eq:comp}
\end{equation}
where $\gamma$ is the adiabatic index, $\epsilon=\delta v_{\|,\textrm{s}}/\delta v_{\perp,\textrm{A}}$ is the ratio of slow to Alfv\'en wave amplitudes, and $\theta_\mathrm{kB}$ is the slow wave propagation angle. The dependence of $\CB$ on $\beta$ is shown in Figure \ref{fig:comp}(c), where it is indeed seen to be linear to within errors of the fit, consistent with Equation (\ref{eq:comp}). There is, however, also much scatter, which may be a result of variation in the other parameters. The slow mode kinetic energy fraction, $\epsilon^2$, can be estimated directly from Equation (\ref{eq:comp}), assuming $\gamma=5/3$ and $\sin(\theta_\mathrm{kB})=1$\footnote{Measurements show that the compressive fluctuations are highly anisotropic at 1\,au \citep{chen12b,chen16b} so that $\sin(\theta_\mathrm{kB})\approx1$.}. The result, as a function of $r$, is shown in Figure \ref{fig:comp}(d), in which it can be seen that $\epsilon^2$ varies with distance to the Sun as $\epsilon^2\propto r^{1.18\pm0.19}$ . This indicates that the lower magnetic compressibility seen by \emph{PSP} near perihelion is not just due to the lower $\beta$ but also a reduced slow mode component, and that there is an additional process acting to increase this compressive component of the solar wind as it travels away from the Sun.

\subsection{Energy Flux and Solar Wind Acceleration}
\label{sec:flux}

\begin{figure}
\includegraphics[width=\columnwidth,trim=0 0 0 0,clip]{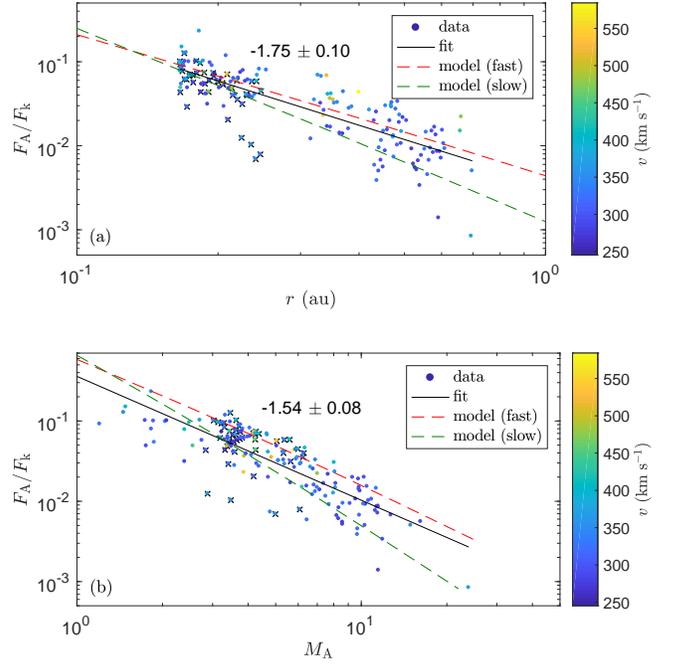}
\caption{(a) Ratio of outward-propagating Alfv\'enic energy flux, $F_\mathrm{A}$, to solar wind bulk kinetic energy flux, $F_\mathrm{k}$, as a function of heliocentric distance, $r$. (b) The same ratio as a function of solar wind radial Alfv\'en Mach number, $M_\mathrm{A}$. In both plots, the black solid line is a power law fit, the red/green dashed lines are the fast/slow solar wind model solutions described in the text, the data points are colored by solar wind speed, $v$, and crosses mark times during connection to the coronal hole in Encounter 1.}
\label{fig:energyflux}
\end{figure}

To determine the role that turbulence plays in the generation of the solar wind, a key measurement is the energy flux of the fluctuations near the Sun. The two dominant contributions to the energy flux in wave- and turbulence-driven solar wind models \citep[e.g.,][]{belcher71b,alazraki71,chandran11} are the enthalpy flux of the outward-propagating Alfv\'enic fluctuations,
\begin{equation}
F_\textrm{A}=\frac{\rho|\delta\mathbf{z}^+|^2}{4}\left(\frac{3}{2}\vrad+\vA\right),
\end{equation}
and the bulk flow kinetic energy flux of the solar wind,
\begin{equation}
F_\textrm{k}=\frac{1}{2}\rho\vrad^3,
\end{equation}
where \vrad\ is the radial component of the solar wind velocity. The ratio of these two terms, calculated from 6-hour rms values of $\delta\mathbf{z}^+$ to capture the full extent of the inertial range and outer scale, is shown as a function of $r$ in Figure \ref{fig:energyflux}(a). It can be seen that this ratio increases as $r$ decreases, and over this range of distances can be fit by a power law $F_\mathrm{A}/F_\mathrm{k}\propto r^{-1.75\pm0.10}$. At 0.17\,au, this ratio is $\sim$20 times larger than at 1\,au, with a value of $\sim$10\%. The Alfvenic flux itself (not shown) also varies as a power law, $F_\mathrm{A}\propto r^{-3.52\pm0.12}$, taking a value $F_\mathrm{A}=0.72$\,mW\,m$^{-2}$ at 0.17\,au. The same ratio, plotted as a function of radial Alfv\'en Mach number, $\MA=\vrad/\vA$, is shown in Figure \ref{fig:energyflux}(b), where it can be seen to take a dependence $F_\mathrm{A}/F_\mathrm{k}\propto\MA^{-1.54\pm0.08}$. Extrapolating this to the Alfv\'en point ($\MA=1$), gives a ratio $\sim$40\%, indicating that within the corona there is likely to be a significant fraction of the solar wind energy flux in Alfv\'enic turbulence.

Also plotted in Figure \ref{fig:energyflux} are the flux ratios from two solutions of the solar wind model of \citet{chandran11}. The model describes a solar wind driven primarily by Alfv\'enic turbulence (that provides both heating and wave pressure), but also contains collisional and collisionless heat fluxes. The solutions are for a fast wind (800 km\,s$^{-1}$ at 1\,au) as described in \citet{chandran11} and a slow wind (337 km\,s$^{-1}$ at 1\,au) chosen to match bulk solar wind values measured during Encounter 1, and the model parameters are given in Table \ref{tab:parameters}. It can be seen that there is a reasonable match between both the fast and slow wind solutions and the measurements, both in terms of the absolute level and approximate power-law trends, indicating that these observations are consistent with such a turbulence-driven solar wind. One consideration is that the observations are for slow solar wind: in Figure \ref{fig:energyflux} all data is for $v<600$ km\,s$^{-1}$. However, near perihelion much of this was Alfv\'enic slow wind, in particular from the equatorial coronal hole during Encounter 1 \citep{bale19,badman19}, to which the wave- and turbulence-driven models are thought to provide a good description \citep{cranmer07}. Times during connection to this coronal hole are marked with crosses in Figure \ref{fig:energyflux}; most of these points lie close to the model solutions, although there are a few significantly below. These correspond to intervals containing quiet radial-field wind during which the turbulent amplitudes are much lower \citep{bale19}.

\begin{deluxetable}{lcc}
\tablecaption{Parameters used in the fast and slow wind model solutions in Figure \ref{fig:energyflux}; see \citet{chandran11} for parameter definitions.\label{tab:parameters}}
\tablehead{\colhead{Parameter} & \colhead{Fast} & \colhead{Slow}}
\tablewidth{\textwidth}
\startdata
$B_\odot$ & 11.8\,G & 10.2\,G\\
$n_\odot$ & $10^8$\,cm$^{-3}$ & $4\times10^8$\,cm$^{-3}$\\
$T_\odot$ & $7\times10^5$\,K & $8.79\times10^5$\,K\\
$\delta v_\odot$ & $41.4$\,km\,s$^{-1}$ & $27.6$\,km\,s$^{-1}$\\
$f_\mathrm{max}$ & 9 & 8\\
$R_1$ & $1.29\,R_\odot$ & $0.3\,R_\odot$\\
$L_{\perp\odot}$ & $10^3$\,km & $10^3$\,km\\
$c_\mathrm{d}$ & 0.75 & 1.35\\
$c_2$ & 0.17 & 0.17\\
$\alpha_\mathrm{H}$ & 0.75 & 0.75\\
$r_\mathrm{H}$ & $5\,R_\odot$  & $30\,R_\odot$\\
\enddata
\end{deluxetable}

\subsection{Power Levels and Inward Fluctuations}
\label{sec:powerlevels}

\begin{figure}
\includegraphics[width=\columnwidth,trim=0 0 0 0,clip]{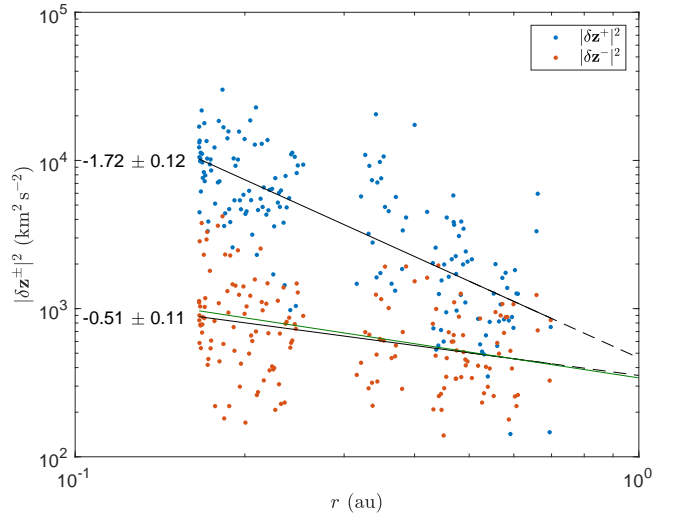}
\caption{Energy in Elsasser fluctuations as a function of heliocentric distance, $r$, with power law fits marked as solid lines. The green line is the predicted $|\dzm|^2$ evolution from Equation (\ref{eq:zminus}).}
\label{fig:radialelsasser}
\end{figure}

It is also of interest to determine the radial variation of the inward-propagating fluctuations to provide information about their origin. Inward-propagating modes are necessary for any nonlinear Alfv\'enic interaction, but any generated inside the Alfv\'en point would not travel further out (since $v<\vA$) meaning that those observed beyond the Alfv\'en point must be generated locally. Figure \ref{fig:radialelsasser} shows the variation of the Elsasser energies calculated over 6-hour intervals as a function of $r$. It can be seen that the inward-propagating fluctuations have a much shallower radial variation than the outward-propagating ones, similar to previous measurements between 0.4 and 3\,au \citep{bavassano00} and qualitatively consistent with predictions from turbulent evolution models \citep[e.g.,][]{verdini07,chandran09b}. The power law variations measured here are $|\dzm|^2\propto r^{-0.51\pm0.11}$ and $|\dzp|^2\propto r^{-1.72\pm0.12}$. 

While several processes may generate inward-propagating fluctuations beyond the Alfv\'en point \citep[see, e.g.,][]{bruno06}, reflection due to the large-scale gradient in \vA\ is thought to be a key mechanism for this, especially at smaller $r$ \citep{heinemann80,dmitruk02,chandran09b,chandran11,perez13,chandran19}. In the model of \citet{chandran11}, the inward-propagating fluctuation amplitude is given by
\begin{equation}
\delta z^-=L_{\perp\odot}\sqrt{\frac{ B_\odot}{B}}\left(\frac{\vrad+\vA}{v_\mathrm{A}}\right)\left|\frac{\partial\vA}{\partial r}\right|,
\label{eq:zminus}
\end{equation}
which describes a balance between its generation by reflection and dissipation through the turbulent cascade. By fitting the measured \vA\ to a power law in $r$, taking the gradient of the fit, and calculating the right-hand-side of Equation (\ref{eq:zminus}) for each data point, the predicted radial variation of $|\dzm|^2$ was determined. The power law fit to the predicted amplitudes, taking $L_{\perp\odot}=1.4\times10^4$\,km as the correlation length and $B_\odot=1.18$\,mT as the magnetic field at the base of the corona, is marked in Figure \ref{fig:radialelsasser} as the green line, and has a variation $\propto r^{-0.58}$. This power law is a good match to that observed, and the values of $L_{\perp\odot}$ and $B_\odot$ are within a reasonable expected range \citep{chandran11,chandran19}, indicating that the inward-propagating fluctuations are consistent with being generated by reflection past the Alfv\'en point. However, it remains possible that other mechanisms such as local driving or parametric decay may also contribute, and further analysis will be needed to test these.

\subsection{Turbulence Outer Scale}
\label{sec:outerscale}

\begin{figure}
\includegraphics[width=\columnwidth,trim=0 0 0 0,clip]{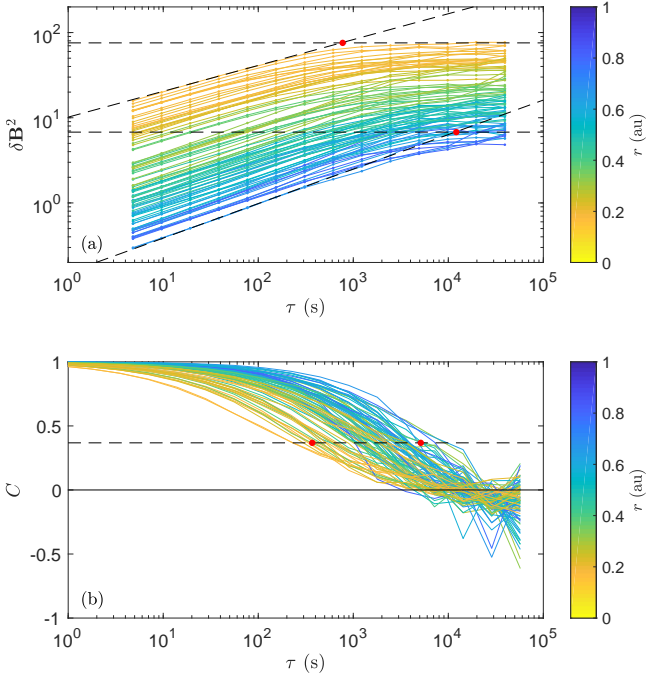}
\caption{(a) 2nd order structure function, $\delta\mathbf{B}^2$, at different heliocentric distances, $r$, with two examples of power-law fits (black dashed) determining the break scale (red dots). (b) Magnetic field correlation function, $C$, at different $r$ showing the correlation time (red dots) at $C=e^{-1}$ (black dashed) for the same two examples.}
\label{fig:sfuncorfun}
\end{figure}

Finally, the evolution of the outer scale of the turbulence was examined. The outer scale can be defined observationally in several ways, e.g., as the beginning of the MHD turbulence scaling range or as the correlation length of the fluctuations. 2nd order structure functions of the magnetic field, 
\begin{equation}
\delta\mathbf{B}^2(\tau)=\left<|\mathbf{B}(t+\tau)-\mathbf{B}(t)|^2\right>,
\end{equation}
can be used to define the scaling ranges, and are shown in Figure \ref{fig:sfuncorfun}(a) calculated at different $r$ over 1-day intervals. It can be seen that they have a steeper scaling at smaller scales and a flat range at larger scales, consistent with the spectra in Figure \ref{fig:bspectrum}\footnote{A flat structure function corresponds to a spectrum with spectral index $-1$ or shallower \citep{monin75}.}. At each distance, a power law fit was made in the inertial range (10 to 100\,s) and the value at large scales was determined by an average of the points with $\tau>10^4$\,s; the break scale, \taub, was determined as the point at which these two lines cross (example fits are shown in Figure \ref{fig:sfuncorfun}(a) for the highest and lowest amplitude curves). Figure \ref{fig:sfuncorfun}(b) shows the normalised magnetic field correlation functions, 
\begin{equation}
C(\tau)=\frac{\left<\delta\mathbf{B}(t+\tau)\cdot\delta\mathbf{B}(t)\right>}{\left<|\delta\mathbf{B}|^2\right>},
\end{equation}
where $\delta\mathbf{B}(t)=\mathbf{B}(t)-\left<\mathbf{B}\right>$, also for 1-day intervals\footnote{While the solar wind correlation time has been shown to depend on the length of the interval used to calculate it \citep{matthaeus82b,isaacs15,krishnajagarlamudi19}, here we choose 1-day intervals as a reasonable compromise, and are more interested in its radial dependence than absolute value.}. The correlation scale, \tauc, can be obtained from $C(\tau)$ in various ways \citep[e.g.,][]{ruiz14,isaacs15}; here it was taken as the point where $C$ decreases such that $C(\tauc)=e^{-1}$.

\begin{figure}
\includegraphics[width=\columnwidth,trim=0 0 0 0,clip]{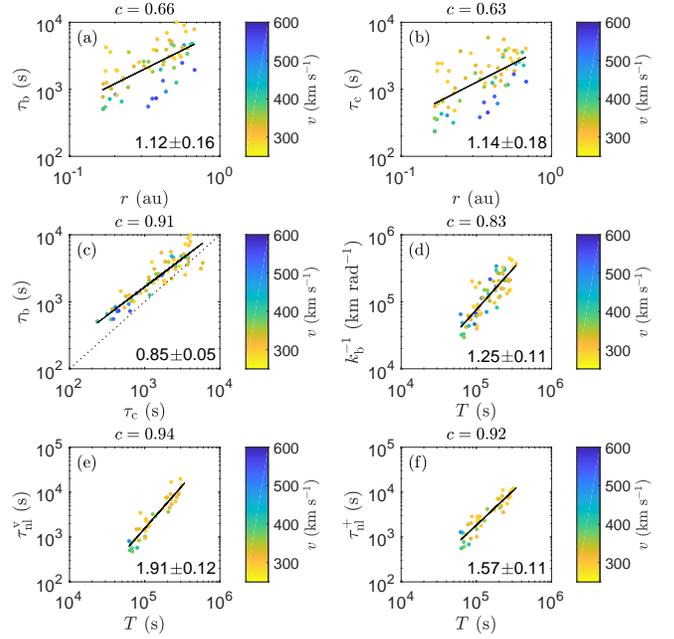}
\caption{(a) Break scale, \taub, as a function of heliocentric distance, $r$. (b) Correlation scale, \tauc, as a function of $r$. (c) Comparison between the two outer scales. (d) \taub\ as a function of travel time from Sun, $T$. (e) Break scale velocity nonlinear time $\tau_\mathrm{nl}^\mathrm{v}$ as a function of $T$. (f) Break scale Elsasser nonlinear time $\tau_\mathrm{nl}^+$ as a function of $T$. For each panel, the correlation coefficient $c$ is given.}
\label{fig:outerscale}
\end{figure}

The radial variation of the two outer scale estimates, \taub\ and \tauc, is shown in Figure \ref{fig:outerscale}(a-b). For both quantities there is a loose positive correlation, and an increase of the outer scale with distance as $\tau_\mathrm{b,c}\propto r^{1.1}$, although there is substantial scatter in the data. Figure \ref{fig:outerscale}(c) shows that there is a good correspondence between \taub\ and \tauc, which is consistent with the structure function and correlation function being directly related quantities \citep{monin75}. It can be seen that much of the scatter in Figure \ref{fig:outerscale}(a-b) can be attributed to the variation in solar wind speed: faster wind has an outer scale at smaller scales. This can be in part because the Taylor shift in faster wind results in the same $k$ appearing at a smaller $\tau$, but can also be due to a physical difference, such as slower wind having a larger travel time and therefore a break scale at larger $\tau$ if it is set by the scale at which the largest eddies have had time to decay \citep{matthaeus86}. Figure \ref{fig:outerscale}(d) shows the Taylor-shifted break wavenumber, $\kb=2\pi/(\taub v)$, as a function of the travel time from the Sun (assuming constant solar wind speed), $T=r/v$, where a better correlation can be seen and the solar wind speed dependence is no longer present. 

Figure \ref{fig:outerscale}(e-f) shows the nonlinear time at the outer scale, defined\footnote{Equating $\tau_\mathrm{nl}^+$ and $(k_\mathrm{b}\delta z^-_\mathrm{rms})^{-1}$ follows from models of inhomogeneous reflection-driven solar wind turbulence, in which the \zp\ fluctuations continually interact with their own reflections and the reflections of the \zp\ fluctuations just ``ahead'' of them, i.e., at larger $r$ \cite[e.g.,][]{velli89,vanballegooijen17}. Note, however, that in some models of imbalanced MHD turbulence without wave reflections, $\tau_\mathrm{nl}^+$ can be much larger than $(k_\mathrm{b}\delta z^-_\mathrm{rms})^{-1}$ \citep[e.g.,][]{beresnyak08}, in which case the nonlinear time for \zp\ would be significantly longer than measured here. Also, these definitions do not include the alignment angle between \dzp\ and \dzm, but since the turbulence here is significantly imbalanced with only small residual energy, this angle is large and the correction is of order unity.} from the velocity fluctuations as $\tau_\mathrm{nl}^\mathrm{v}=(k_\mathrm{b}\delta v_\mathrm{rms})^{-1}$ and from the \dzm\ fluctuations (i.e., the timescale for the \dzp\ fluctuations) as $\tau_\mathrm{nl}^+=(k_\mathrm{b}\delta z^-_\mathrm{rms})^{-1}$, as a function of $T$. It can be seen that in both cases there is a good correlation, but the power-law dependence is much stronger than linear and $\tau_\mathrm{nl}^\mathrm{v},\tau_\mathrm{nl}^+\ll T$. This indicates that the fluctuations in the large-scale flat scaling range have had significant time for nonlinear processing, and increasingly so closer to the Sun. This would suggest that this range might not be a simple spectrum of non-interacting waves, but could be undergoing nonlinear interactions, as in more recent models of the $1/f$ range \citep{velli89,verdini12,perez13,chandran18,matteini18,matteini19}.

Finally, it is of interest to compare the correlation times of the different  Alfv\'enic turbulence fields, which are shown in Table \ref{tab:cor} for the day of Perihelion 1, 6th November 2018. The magnetic, velocity, and outward Elsasser fields have correlation times of $\tauc\sim$ 7\,min, whereas the inward Elsasser field has a correlation time 8 times longer. All correlation times are shorter than those seen at 0.3\,au by \emph{Helios} \citep[][Table 1]{tu95}, consistent with the radial trend in \tauc\ described above. In addition, the ratio between \zm\ and \zp\ correlation times is greater than seen by \emph{Helios} at 0.3\,au. The observation that \zm\ has a longer spacecraft-frame correlation time than \zp\ is consistent with models in which reflection of \zp\ fluctuations is the source of the \zm\ fluctuations. Since \zp\ fluctuations reflect more efficiently at lower frequencies \citep{heinemann80,velli93}, the energy-weighted average frequency of the \zm\ fluctuations is smaller than that of the \zp\ fluctuations, which implies that the characteristic correlation length of the \zm\ fluctuations is larger than that of the \zp\ fluctuations. Therefore, the observed difference in correlation times is consistent with the interpretation of reflection-generated inward-propagating fluctuations discussed in Section \ref{sec:powerlevels}.

\begin{deluxetable}{lc}
\tablecaption{Measured correlation times for the different MHD turbulence fields.\label{tab:cor}}
\tablewidth{\columnwidth}
\tablehead{\colhead{Field} & \colhead{\tauc}}
\startdata
$\mathbf{B}$ & 417\,s\\
$\mathbf{v}$ & 419\,s\\
$\mathbf{z}^+$ & 407\,s\\
$\mathbf{z}^-$ & 3,300\,s\\
\enddata
\end{deluxetable}

\section{Discussion}
\label{sec:discussion}

In this paper, the properties of solar wind turbulence were measured in situ down to a heliocentric distance of 0.17\,au for the first time. While many of the measured properties are shared with measurements nearer 1\,au, significant differences include increased power levels (by more than two orders of magnitude in magnetic fluctuations, and one order of magnitude in total energy), a $-3/2$ spectral index in all fields, a significantly smaller compressive component of the turbulence, a much smaller outer scale at which the nonlinear time is less than the travel time from the Sun, and an increase in the turbulence imbalance (measured through the cross-helicity \sigmac\ or Elsasser ratio \rE) that is consistent with generation of the inward-propagating component by reflection. The energy (enthalpy) flux in the turbulence increases to a significant fraction of the bulk solar wind kinetic energy flux in a manner consistent with models in which the solar wind is driven by this flux.

The Alfv\'enic turbulence spectra presented here were measured closer to the Sun and at higher frequencies than has previously been possible, e.g., with \emph{Helios} \citep{tu95,bruno13}. The spectra at 0.17\,au (Figures \ref{fig:bspectrum}--\ref{fig:allspectra}) have inertial range spectral indices of $\alpha\approx -3/2$ for both inward and outward-propagating fluctuations. These spectra are similar to, although a little flatter than, the spectra predicted by \citet{lithwick07}, whose model relies upon assumptions that may also describe reflection-driven turbulence in the solar wind (see \citet{chandran19} for a more detailed discussion of this point). The reason that the observed spectra are flatter than the $-5/3$ spectra predicted by \citet{lithwick07} might be the presence of scale-dependent dynamic alignment \citep{boldyrev06} or intermittency \citep{chandran15,mallet17}, both of which progressively weaken the nonlinearity in a critically balanced cascade. The $\alpha\approx -3/2$ spectra are also consistent with some models of homogeneous imbalanced MHD turbulence that do not invoke wave reflection \citep{perez09,podesta10d}. Another possibility for the spectral index trend in Figure \ref{fig:bindex} is the turbulence transitioning from a much shallower spectrum closer to the Sun, e.g., the reflection-driven cascade model of \citet{velli89}, which predicts a $k^{-1}$ spectrum. A further possibility is that at smaller $r$ there is a more significant effect of the driving, which may effect the spectrum in different ways. Firstly, the properties of the turbulent cascade may differ depending on whether it is forced at large scales or decaying; \citet{chen11a} found that in a simulation of Alfv\'enic turbulence the spectral indices of all fields vary from $-3/2$ to $-5/3$ as the simulation transitions from a forced to a decaying state. Secondly, closer to the Sun there may be a stronger signature of the driving itself throughout the spectrum (as discussed later in this section). Future orbits of \emph{PSP} at smaller $r$ will hopefully allow these various possibilities to be distinguished.

The decrease in magnetic compressibility closer to the Sun was shown to be associated with both a decrease in $\beta$ and a reduction in the kinetic energy in the slow mode component of the turbulence (Figure \ref{fig:comp}). One possible reason for the slow mode component to increase as the solar wind travels from the Sun is continual local driving, e.g., from velocity shears \citep{roberts92} or parametric decay  \citep{delzanna01,tenerani13,bowen18b}. However, it is also possible that another processes is acting to suppress the fluctuations in $|\mathbf{B}|$ nearer the Sun. It has been proposed that a higher-order effect of large-amplitude Alfv\'en waves is to reduce the variations in $|\mathbf{B}|$ which can be thought of an effect of the magnetic pressure force \citep{cohen74,vasquez96}, similarly to the effect of the pressure anisotropy force at high $\beta$ found recently \citep{squire19}. Future work could include further investigation of these possibilities and the nature of the compressive component.

The increase of the outer scale with $r$, approximately as $\propto r^{1.1}$, is qualitatively consistent with previous studies at larger distances. Specifically, the variation is consistent with previous results for the $1/f$ break evolution between 1.5 and 5\,au in polar fast wind \citep{horbury96a} although shallower than the variation found in ecliptic fast wind from 0.3 to 5\,au \citep{bruno13} and steeper than found for the radial variation of the correlation scale from 0.3 to 5\,au \citep{ruiz14}. The finding of the nonlinear time at the break scale being much less than the travel time from the Sun would indicate that the fluctuations in the flat scaling range (larger than the break scale) have had significant time for nonlinear processing, raising the question of why the break is not at lower frequencies. Possibilities for this include a nonlinear cascade that produces a $1/f$ spectrum \citep{velli89,verdini12,perez13,chandran18} or that the fluctuations in this range have reached a saturated state and cannot grow to larger amplitudes \citep{villante80,matteini18,matteini19}. The radial variation of the outer scale is the same (to within errors) as that of the ion break scale \citep{duan19} indicating that the width of the MHD inertial range, $\sim$3 decades, stays approximately constant from 0.17 to 1\,au. This has important implications, e.g., the level of anisotropy at kinetic scales is determined by the extent of the inertial range \citep{goldreich95} and the possible heating mechanisms there depend on the level of anisotropy \citep{schekochihin09}.

The increase of energy flux in the fluctuations near the Sun, compared to the bulk solar wind kinetic energy flux, was found to be consistent with solutions of the turbulence driven solar wind model of \citet{chandran11} down to 0.17\,au. The enthalpy flux in the outward-propagating fluctuations (\dzp) was found to be $\sim$10\% of that in the bulk kinetic energy at this distance and $\sim$40\% if extrapolated to the Alfv\'en point, indicating a significant turbulence flux is likely within the corona. This increase of Alfv\'enic flux towards the Sun is also consistent with remote observations of Alfv\'en waves in the chromosphere and corona, which were measured to contain sufficient energy to accelerate the fast solar wind \citep{depontieu07,mcintosh11}. The \emph{PSP} results indicate that turbulence-driven models \citep[e.g.,][]{cranmer07,verdini09,chandran09b,chandran11,vanderholst14} remain a viable explanation for the acceleration of the solar wind from open field regions.

The radial variation of the inward-propagating component (\dzm) was also found to be consistent with the reflection-driven model of \citet{chandran11}, making the reflection of the outward-propagating fluctuations (from the large-scale gradient in \vA) to form the inward ones a viable explanation for the decrease in imbalance at larger distances (Figure \ref{fig:sigma} and \ref{fig:radialelsasser}). This trend is qualitatively consistent with previous measurements at larger radial distances \citep{roberts87a,tu95,bavassano98,bavassano00,matthaeus04,breech05}. However, it is possible that other mechanisms, such as local driving \citep{roberts92,matthaeus04,breech05} and parametric decay \citep{marsch93c,delzanna01,tenerani13,bowen18b} may also contribute, and future observations will be needed to distinguish these possibilities.

One relevant question is the relation between the turbulence and the large amplitude fluctuations known as ``switchbacks'', ``jets'', or ``spikes'', that appear more prominent closer to the Sun \citep{bale19,kasper19,horbury19,dudokdewit19,mcmanus19}. These are Alfv\'enic fluctuations which significantly change the magnetic field direction, and appear to occur in patches \citep{horbury19} with quiet periods in between \citep{bale19} and are correlated and have a scale-invariant distribution \citep{dudokdewit19}. The origin and role of these structures is an open question, in particular whether they are generated by the turbulence, are not initially but then become part of the cascade, or are unrelated altogether. Initial analysis indicates that while the amplitude of the fluctuations is lower in the quiet periods \citep{bale19} and various kinetic waves become detectable \citep{bowen19,malaspina19}, in the inertial range both types of wind have a $-3/2$ spectrum consistent with turbulence, although the extent of this might be smaller in the quiet periods \citep{dudokdewit19}. One possible interpretation is that these large-amplitude fluctuations represent the remnant of driving processes at the Sun that become part of the turbulent cascade as the solar wind expands. In this paper, all fluctuations are considered part of the turbulence cascade, although future work could investigate this relationship further.

With future \emph{PSP} orbits, it will be possible to see how the trends measured in this paper continue to smaller distances to provide more insight into the fundamental nature of the cascade, directly measure the turbulence energy flux within the solar corona to determine its contribution to solar wind acceleration, examine the turbulence \citep{alexandrova13a,chen16b,chen17,duan19} and field-particle interactions \citep{chen19} at kinetic scales to understand how it heats the corona and inner solar wind, and perhaps probe the nature of the turbulence driving mechanisms. Such data, closer to the Sun and within the Alfv\'en point, promises to continue revealing more about the nature of plasma turbulence and the role it plays in the near-Sun environment. 

\section*{Acknowledgements}

CHKC is supported by STFC Ernest Rutherford Fellowship ST/N003748/2. SDB acknowledges the support of the Leverhulme Trust Visiting Professorship program. DB was supported by STFC grant ST/P000622/1. BDGC was supported in part by NASA grants NNX17AI18G and 80NSSC19K0829. KGK is supported by NASA ECIP Grant 80NSSC19K0912. We thank the members of the FIELDS/SWEAP teams and \emph{PSP} community for helpful discussions. \emph{PSP} data is available at the SPDF ({https://spdf.gsfc.nasa.gov}).

\bibliography{bibliography}

\end{document}